\documentclass[aps,prb,twocolumn,showpacs,graphix]{revtex4}
\usepackage{graphicx}

\begin{document}

\title{Periodic Density of States Modulations in Superconducting $\rm Bi_2Sr_2CaCu_2O_{8+\delta}$}

\author{C. Howald$^1$, H. Eisaki$^2$,  N. Kaneko$^3$, M. Greven$^{2,3}$, and A. Kapitulnik$^{1,2}$ }

\affiliation{$^1$Department of Physics, Stanford University, Stanford, CA 94305, USA\\
$^2$Department of Applied Physics, Stanford University, Stanford, CA 94305, USA\\
$^3$ Stanford Synchrotron Radiation Laboratory, Stanford, CA 94309, USA\\ }

\date{\today}

\begin{abstract}

In this paper we show, using scanning tunneling spectroscopy
(STS), the existence of static striped density of electronic
states in nearly optimally doped $\rm Bi_2Sr_2CaCu_2O_{8+\delta}$
in zero field. The observed modulation is strongest at roughly
half the superconducting gap energy and is aligned with the Cu-O
bonds, with a periodicity of four lattice constants, exhibiting
features characteristic of a two-dimensional system of line
objects. These features also exhibit asymmetries between the two
orthogonal directions locally, pointing to a possible broken
symmetry state (i. e., stripe phase). We further show that the
density of states modulation manifests itself as a shift of
states from above to below the superconducting gap. The fact that
a single energy scale (i. e., the gap) appears for both
superconductivity and stripes suggests that these two effects
have the same microscopic origin.

\end{abstract}
 \maketitle

\section{INTRODUCTION}
One of the most important problems for condensed matter physics is
the determination of the ground states of strongly correlated
electron systems. In particular, high-temperature superconductors
(HTSC) appear to have unusual ground states because of strong
electronic correlations. Theoretical
\cite{zaanen,ek1,ek2,polkovnikov,zaanen1,white} and experimental
\cite{tranquada,lake,lake1,yslee,mitrovic,hoffman,khaykovich,zhou}
evidence has been mounting in support of the possibility that
their ground state exhibits spin and charge density waves (SDW
and CDW), which are primarily one-dimensional (i. e., stripes)
and which either compete with or promote superconductivity.
Coexistence of CDW or SDW and superconductivity has previously
been reported in the lower $T_{c}$ materials
\cite{tranquada,lake,lake1,khaykovich,zhou,yslee} and in the
presence of large magnetic fields \cite{mitrovic,hoffman}.

In inelastic neutron scattering experiments on $\rm
La_{2-x}Sr_xCuO_4$ near optimum doping (x=0.163), and in the
presence of a magnetic field,  Lake {\it et al.}
\cite{lake,lake1} found strong scattering peaks at the four
$k$-space points: ($\pi/a_0$)[(1/2,1/2) $\pm \delta$(0,1/2)] and
($\pi/a_0$)[(1/2,1/2) $\pm \delta$(1/2,0)], where $\delta \sim
0.25$ and $a_0$ is the lattice constant. This implies a
spin-density of periodicity $\sim 8a_0$, which the authors find
extends to more than 50$a_0$, much beyond the vortex core.  Also,
Mitrovic {\it et al.} \cite{mitrovic} reported similar
periodicity in a high-field NMR imaging experiment. Using STM,
Hoffman {\it et al.} \cite{hoffman} reported scanning tunneling
spectroscopy on $\rm Bi_2Sr_2CaCu_2O_{8+\delta}$ single crystals
revealing a ÔcheckerboardÕ of quasiparticle states with four
unit-cell periodicity surrounding vortex cores. In $k$-space this
corresponds to Bragg peaks in the local density of states at
(2$\pi /a_0$)($\pm$1/4,0) and (2$\pi /a_0$)(0,$\pm$1/4). This
structure was found around the energy ($\sim$7 meV) of a feature
previously observed in the spectra of vortices in the same field
range \cite{renner2,pan2}. The ÔcheckerboardÕ structure extends
over $\sim 20a_0$, again indicating ordering outside the vortex
cores \cite{demler}.

The absence of data showing stripes in the higher $T_c$ materials
in zero field has lent support to the idea that stripes are
competing with high temperature superconductivity. However, we
showed\cite{howald}, using scanning tunneling spectroscopy (STS),
a first indication of the existence of static striped density of
electronic states in nearly optimally doped $\rm
Bi_2Sr_2CaCu_2O_{8+\delta}$ in zero field. This modulation is
aligned with the Cu-O bonds, with a periodicity of four lattice
constants, and exhibits features characteristic of a
two-dimensional system of line objects. We further showed that
the density of states modulation manifests itself as a shift of
states from above to below the superconducting gap. The fact that
a single energy scale (the gap) appears for both
superconductivity and stripes suggests that these two effects
have the same origin. In the present paper we expand on our
initial results, studying further the energy dependence of the
modulation and its interrelation with quasiparticle scattering.
Our main result in the present paper is that the four-period
peaks in the Fourier transform of the data are present at all
energies, including very low energies. At the same time, peaks at
other $k$-vectors appear at higher energies, suggesting the
contribution of quasiparticle-like scattering from impurities, as
suggested by Wang and Lee \cite{wanglee}. The two effects are
separable, suggesting that a simple model of two-dimensional band
electrons cannot explain all the observed physics, corroborating
the photoemission findings on the same system
\cite{bogdanov,bogdanov1,kaminski}.

\section{EXPERIMENTAL}
STS allows one to measure, on an atomic scale, the electronic
density of states. This measurement of the local density of
states (LDOS) makes STS a powerful tool for investigating
correlated electron systems. In this study we used $\rm
Bi_2Sr_2CaCu_2O_{8+\delta}$ because it has a high $T_c$($\sim$90
K), and since it cleaves easily, yielding large, stable,
atomically flat surfaces. The single crystals of $\rm
Bi_2Sr_2CaCu_2O_{8+\delta}$ used in this study, grown by a
floating-zone method, were annealed to be slightly overdoped,
yielding a superconducting transition temperature of 86 K. The
samples were cleaved (between the BiO planes) at room temperature
in a vacuum of better than $1 \times 10^{-9}$ torr, then
transferred in less than one minute to the low-temperature STM,
where cryopumping yields orders of magnitude better vacuum. Data
were taken at 8 K with an iridium tip at a sample bias of -200 mV
and a setpoint current of -100 pA. The choice of setpoint
establishes the relatively arbitrary normalization condition for
the raw differential conductance (dI/dV), which is proportional to
the local density of states. The data sets used in this paper
consist of $64 \times 64$ pixel images with dI/dV spectra taken at
each point.  In addition, we took $256 \times 256$ pixel
topographic scans over the same areas. Fig.\ 1a shows the
topography of a typical area: the surface exhibits clear atomic
resolution. In particular, the superstructure in the BiO plane
\cite{kirk} with average periodicity $\sim$27 \AA (as well as the
location of the individual Bi atoms) is evident in all scans and
provides a reference direction for our study. As reported earlier
\cite{howald1,pan1,lang}, all samples also exhibit strong
inhomogeneity in the local spectroscopy. This manifests itself
here as large variation in the superconducting gap size
($\Delta$), as measured by the location of the peak in the local
density of states (Fig.\ 1b). Overall the spectra evolve from
small superconducting gaps ($\Delta \sim 30$ mV) to very large
gaps ($\Delta > 60$ mV) with much reduced coherence peaks
reminiscent of the pseudogap \cite{loeser,renner1,howald1}.

\section{RESULTS AND ANALYSIS}

\subsection{Fourier Analysis}

Fig.\ 2a shows a map of the LDOS for 15 mV sample bias.  While
there are some features in the real-space image, they are more
readily observed in Fourier space \cite{sprunger}. Therefore, we
will focus for the moment on Fig.\ 2b, which shows the power
spectrum, namely the amplitude squared of the Fourier transform,
of the data in Fig.\ 2a. Fig.\ 2b exhibits a large amount of
weight at small wavevectors, coming mostly from the random
contributions of the inhomogeneous gap. In addition, there are
two bright peaks appearing along the $k_x$ direction which
originate from the superstructure. Finally, Fig.\ 2b clearly
shows four peaks, oriented 45 degrees to the superstructure and
at points corresponding to, within experimental uncertainty,
periodicities of ($\pm 4a_0$, 0) and (0, $\pm 4a_0$), oriented
along the copper-oxygen bonds. The real-space data (Fig.\ 2a)
show a two-dimensional pattern in the LDOS that is oriented along
the diagonals; however, as is evident from Fig.\ 2b, contributions
from features at other wavelengths obscure this modulation. The
period of the modulation is shown clearly in Fig.\ 3, the
autocorrelation function of Fig.\ 2a. This figure has been rotated
approximately 45 degrees clockwise from Fig.\ 2a to place the
copper-oxygen-copper directions horizontally and vertically. The
maxima are separated by $\sim 4a_0$.  The apparent rhombic
distortion of the central nine maxima comes from the near
coincidence of the upper left and low right corners with the
superstructure period, which moves these maxima to slightly
larger distances.

In order to further elucidate the spatial structure of the
observed modulation, we can filter from Fig.\ 2b the
contributions to the LDOS modulation that are far away from the
(2$\pi /4a_0$)($\pm$1,0) and (2$\pi /4a_0$)(0,$\pm$1) points. In
general, filtering in reciprocal space is performed by multiplying
the complex two-dimensional Fourier transform $F(k_x,k_y)$ of the
real space image $f(x,y)$ with a real filter mask $m(k_x,k_y)$
and subsequently calculating the inverse transform. The filter
mask must be a symmetric function with respect to {\bf k}=0.  In
order to avoid artifacts, one should take care that there is
little weight around the edges of the filter, and that the edges
of the filter are not particularly sharp\cite{briner}. In
addition, examination of the resulting image for a variety of
widths can show whether artifacts are important since the
wavevector of such artifacts will depend on the sizes of the
filter. For our data, we have chosen a radial Gaussian mask
function with radius $\sigma$ centered around the four-period
points. Obviously such a procedure will pick out a preferred
wavelength; its validity here rests on the observation of
relatively large weight in these regions in the raw power
spectrum, and the relatively small weight at the filter edges.
This procedure is hardly novel: similar Fourier filtering was
successfully used to remove the topographic background in order
to enhance the Friedel oscillations \cite{friedel} of
Be(10$\bar{1}$0) STM images \cite{briner}, for example.

Fig.\ 4 shows the real-space data after such Fourier filtering
with a standard Gaussian filter function using $\sigma$= (2$\pi
/15a_0$). Varying the width of this filter has no qualitative
effect on the resulting image, and the size of the above filter
implies about 30 pixels within the FWHM around the center of each
Bragg-spot. The modulation partially visible in Fig.\ 2b is now
clear: Fig.\ 4 shows a dominant four-period modulation of the
density of states that is almost checkerboard like, but with
defects characteristic of a two-dimensional four-fold structure.
In particular we observe dislocations in the form of extra
half-rows. These defects locally increase or decrease the
periodicity. From the separation of the defects in this map we
estimate the correlation length to be $\sim20a_0$.

The defects noted above in Fig.\ 4 can also explain the weight
distribution of the four-period peak in the Fourier-transform.
Given the approximately 80 \AA  correlation length of the
modulation, we would expect a peak width of about $2\pi/20a_0$.
However, because the transform is taken over an area that is only
about two correlation lengths wide, we expect to see contributions
at only some of the wavevectors in this region.  Since any
dislocation alters the local periodicity of the checkerboard, we
see a number of peaks at positions near $2\pi/4a_0$.  Assuming
that these reflect a single peak broadened by the finite
correlation length, we therefore place the peak at $2\pi/4a_0 \pm
2\pi/40a_0$.

\subsection{Energy Dependence}

To further understand the nature of the modulation near
$k_{x,y}=2\pi / 4a_0$, we studied the energy dependence of the
effect.  Fig.\ 5 shows the power spectra of the conductance over
this area for a number of voltages. As above, the axes have been
rotated to place the Cu-O-Cu directions parallel to the axes. For
all sub-gap energies, there is significant weight in the vicinity
of the $2\pi/4a_0$ points, as shown by the circles. However, the
small wavevector (long wavelength) contributions increase with
increasing energy, up to the nominal gap size, as we would expect
from the spatial variation in gap size.  This means that the
$2\pi/4a_0$ points are most relatively prominent for lower
energies.

The four-period peaks are not the only features observed in the
power spectra. Also evident are the superstructure points at $
\sim 2\pi/a_0 (\pm 0.1,\mp 0.1)$.  In addition, for most energies
one can see faint second-order superstructure spots. Curiously,
there also appear spots perpendicular to the superstructure at
the points $ \sim 2\pi/a_0 (\pm 0.1,\pm 0.1)$, which do not appear
to disperse with energy.  Finally, while there is a significant
weight at many wave vectors within the box bounded by the
$k_{x,y}=2\pi / 4a_0$ points, there is, with the exception of the
second order superstructure spots, essentially no weight outside
of it.

Fig.\ 6 shows a number of line cuts along the $(\pi,0)$ and
$(0,\pi)$ directions.  Each curve shows the magnitude of the
Fourier transform along the Cu-O-Cu direction for a different
energy. The curves have been computed by Gaussian averaging in
the vicinity of the desired point with a FWHM of one pixel width
to remove pixel size artifacts and account for the (small)
uncertainty in knowing the Cu-O-Cu direction. The peaks around
$k_{x,y}=2\pi / 4a_0$ show up at all these sub-gap energies in
both biases (although only the positive bias is shown here for
clarity). While there appears to be a small change, from $0.25
\times (2\pi /a_0)$ at low energy to about $0.23 \times (2\pi
/a_0)$ around 30 mV, no significant dispersion is found for these
peaks. Whether this small shift comes from the appearance of an
additional dispersive or non-dispersive feature, an increase in
the low-wavevector background, or a small dispersion in these
$k_{x,y} \sim 2\pi / 4a_0$ points should be resolved by
additional data.

In a recent theoretical paper, Wang and Lee \cite{wanglee}
explored the possibility that band-structure quasiparticles will
exhibit quantum interference due to the presence of point defects
in the system. This would manifest itself as strongly dispersing
features in the Fourier transform of the LDOS as a function of
energy. Clearly, such an effect will be strong at high energies
(bias voltages) and will disappear close to zero bias due to the
lack of phase space for the possible $k$'s. Indeed, Wang and Lee
find essentially no peaks at any $k$ for energies close to zero
bias in their theoretical study. Since we see a feature at $0.25
\times (2\pi /a_0)$ for $V=0$ mV and it shows no dispersion up to
20 mV, it is clear that we observe something other than the
dispersion predicted by Wang and Lee. We will come back to this
point in the next section.

\subsection{Phase Information}

In addition to the information extracted above from the magnitude
of the Fourier transform, there is also information contained in
the phase for the variety of energies sampled. We now use this to
further understand the four-period modulation observed and
contrast it with other features discussed above (Figs.\ 5,6).
Fig.\ 7 shows that the modulation in the LDOS exhibits a strong
energy dependence. Here we show the value of the Fourier
transform at the two independent peaks from Fig.\ 2b as a
function of the bias voltage. The error bars are determined by
modeling the variations in dI/dV(x,y) at each energy by
uncorrelated noise of the same amplitude. The arbitrary phase
(dependent on the point about which the Fourier transform is
computed) has been set to maximize the real part, plotted in red,
which changes sign at about $\pm$40 mV. The variation in the
imaginary part of the signal is considerably smaller, consistent
with zero within the uncertainty. This is important, since it
shows that the location in real-space of the density of states
(DOS) modulation is the same at all measured energies, although
at about $\pm$40 mV the positions of the maxima and the minima
switch.

The constancy of the phase can be illustrated in another manner,
reminiscent of the analyses of neutron scattering results: Fig.\ 8
is similar to Fig. 5, but integrated in energy. Each panel shows
the power spectrum of the current at that energy, which is equal
to the integral from zero to V of dI/dV. It is clear that with
increasing energy, the four-period modulation remains relatively
strong, while many other features fade away. The features that
remain at many energies are apparently summed with the same
wavevector and phase, i. e., they are a non-dispersive and pinned.
On the other hand, many of the other features (internal to the
four spots in $k$-space) are reduced in intensity, indicating more
variation in either phase or $k$-space location with energy.  To
focus on the peaks at $k_{x,y}=2\pi / 4a_0$, we show in Fig.\ 9
the contrast between the raw dI/dV and the integrated dI/dV for
line scans along the $(0,0)$ to $(\pi,0)$ direction as a function
of wave-vector and energy. The fact that the peak at $k_{x}=2\pi
/ 4a_0$ is enhanced as a function of energy for the integrated
curves reflects the non-dispersing nature of this modulation
\cite{kivfradkin}.

Furthermore, the energy dependence of the Fourier transform
(Fig.\ 7) shows that relative spectral weight is shifted to
sub-gap energies (with a peak at $\pm$25 mV) from intermediate
energies (with weight between 50 and 150 mV). The comparison of
the above result with the energy of the superconducting gap
($\Delta \sim$ 40 mV) provides strong circumstantial evidence
that the striped quasiparticle density and superconductivity are
intimately connected \cite{demler1}. Since the integrated density
of states from zero to infinity is the total charge, the shift of
weight from intermediate to low energies coupled with the small
magnitude of the Fourier components ($< 5 \%$ of the LDOS)
indicates that the total charge participating in this density of
states modulation is very small, as is expected for itinerant
systems with strong Coulomb interactions
\cite{castellani,perali,kivaeppli}. On the other hand, it is
important to note that energy independent spatial variations in
the DOS would not appear in this analysis because the
normalization is set by maintaining constant current (which is
proportional to the integrated DOS from zero to the setpoint
voltage). We are sensitive to features in the shape of the LDOS,
not the overall magnitude. Energy independent changes would
however appear in the so-called topographic signal (Fig.\ 1a),
which actually maps a contour of constant integrated density of
states. The power spectrum of the topograph reveals no such peaks
above the noise; however, this is inconclusive because of the
small amount of total charge expected to participate in the
modulation for this system\cite{castellani,perali,kivaeppli}.

\subsection{Anisotropy}

To examine whether the LDOS modulation in our experiments is
anisotropic, we performed Fourier analysis on many smaller
sub-regions of a slightly larger area (Fig.\ 10).  In Fig.\ 10c
and 10d we show maps of the spatial variation of the amplitude of
the two perpendicular pairs of Bragg peaks. The color at each
point is determined by taking the magnitude of the Fourier
transform at the four period point for the $32 \times 32$ pixel
dI/dV(15mV) map centered at this point.  In addition to the small
scale variations which come from noise and finite size effects,
it is clear that each varies on the scale of roughly 100 \AA. In
addition, these large scale features exhibit significant
anticorrelations between the two maps: there are several regions
in space where one map shows a local maximum while the other
shows a local minimum. This indicates that locally there is a
broken symmetry, and that the underlying order is not
two-dimensional, but one-dimensional. The two-dimensional nature
of the patterns observed appears to come from the
interpenetration of these perpendicularly oriented stripes.

\section{DISCUSSION}

``Stripe order'' usually describes unidirectional density wave
states, which can involve unidirectional charge modulations
(``charge stripes'') or coexisting charge and spin density order
 (``spin stripes'').  Such a state in HTSC, with spatial segregation
of charge, was first suggested by Zaanen and Gunnarson
\cite{zaanen} and by Schulz \cite{schulz}.  Stripe order is
expected to occur via a quantum phase transition (i. e. at $T=0$),
where the broken symmetry state is one that breaks rotational
symmetry and translation symmetry perpendicular to the stripe
direction (in a crystal we mean breaking the crystal symmetry
group.)  For HTSC, in the regime where superconductivity is
present, Kivelson and Emery \cite{ek1} argued that while no long
range stripe order will be present for $T>0$, the proximity of
the system to the stripe quantum critical point implies strong
stripe fluctuations.  In this stripe-liquid phase stripes are
dynamical objects and thus cannot be detected with static probes
such as STM. However, even a small amount of quenched disorder
will pin the stripes, hence revealing the nature of the nearby
ordered stripe phase \cite{ek1,polkovnikov,hasselmann}.
Vortex-induced pinning of stripes was used by Hoffman {\it et al.}
\cite{hoffman} to explain the observed density modulation near
vortex cores. Random point defects, present in all HTSC samples,
could play a similar role, thus explaining the origin of the
stripe patterns we observe.

While the analysis presented above concentrated on the robust
feature at $k$-vectors $2\pi/4a_0$ along the Cu-O bonds, it is
evident that our data show more than just that modulation.
Features, which may be dispersing similar to the suggestion of
Wang and Lee \cite{wanglee}, appear in the data.  Thus, while it
is clear that the simple picture of Wang and Lee cannot explain
our data, the extra features suggest that the stripe picture
alone is not comprehensive. Explaining the electronic structure
data using a simple band structure approach was already shown to
be inadequate by recent ARPES studies
\cite{loeser,bogdanov,feng,orgad,bogdanov1,kaminski}. In the
normal state (i. e. at temperatures above $T_c$), the main
problem arises in analyzing the single hole spectral function as
a function of frequency and wavevector. While for a fixed {\bf k}
the spectral function showed only broad features as a function of
$\omega$, the structure of the spectral function became sharp
when measured at a fixed frequency $\omega$, and as a function of
{\bf k}. This effect points to the absence of stable excitations
with the quantum numbers of the electron, hence, a non-Fermi
liquid situation exists in the normal state of the cuprates. In
the ordered state (the superconducting state) quasiparticles seem
to become well defined \cite{feng1}. This fact was taken at face
value by Wang and Lee to suggest that in the superconducting
state they should be able to use the simple band structure
approach.

However, the same ARPES data clearly indicate that the
superconducting state is anomalous as well, with a coherence-like
peak that tracks the superfluid density and the hole doping
instead of the gap \cite{feng1}, and an anomalous nodal
quasiparticle lifetime \cite{valla}. Here it was found that the
inverse lifetime of the excitations is proportional to
temperature, independent of binding energy, for low energies, and
is proportional to energy, independent of temperature for large
binding energies. This ``marginal Fermi liquid" behavior
\cite{varma} is unaffected by the occurrence of superconductivity
as the temperature is lowered. Nevertheless oscillations that are
of a single particle nature will occur even when there are no
well defined quasiparticles, so long as there are some elementary
excitation in the system with a well defined dispersion
relation.  If we take the underlying band structure measured by
ARPES for $\rm Bi_2Sr_2CaCu_2O_{8+\delta}$ at similar doping as a
guideline\cite{bogdanov1,kaminski}, we find discrepancies with the
quasiparticle scattering interference picture in two regimes.
First, as we noted above, we observe Fourier peaks at $0.25
\times (2\pi /a_0)$ at and near zero bias.  It is easy to see
that the quasiparticle interference picture cannot produce such
peaks at low energies \cite{kivelson,wanglee}. Assuming
quasiparticles with $k$-dependent energy: $E_k=\sqrt{\Delta_k^2 +
\epsilon_k^2}$, where $\Delta_k={{\Delta_0} \over {2}}[cos(k_xa_0)
- cos(k_ya_0)]$ is the d-wave superconducting gap and
$\epsilon_k$ is the normal-state quasiparticle dispersion
(assuming well-defined quasiparticles exist) measured from the
Fermi surface. We assume scattering on the Fermi surface from
state ${\bf k}$ to ${\bf k^\prime}$ with ${\bf q}={\bf
k^\prime}-{\bf k}$, such that ${\bf q}=(2\pi/4a_0, 0)$. In that
case ${\bf k}=(-\pi/4a_0, k_y)$, and ${\bf k^\prime}=(\pi/4a_0,
k_y)$.  If we extract $k_y$ directly from the ARPES data
\cite{bogdanov1,kaminski}, we find $k_y \approx 0.6 (\pi/a_0)$.
For $\Delta_0 \sim 30$ meV \cite{loeser,ding1,fedorov}, which is
also the minimum gap found on this type of samples using STM
\cite{howald1,lang}, this gives an estimate of the lowest energy
for quasiparticle scattering at this wavevector of approximately
$\Delta_0/2 \sim 15$ meV.  The correct energy cutoff may be
somewhat different depending on the exact details of the
band-structure, but this analysis certainly excludes the energies
around zero bias. Second, as noted above, the assumption of
quasiparticles with well defined dispersion should cease to be
justified when the energy is $\sim k_BT_c \sim 15$ meV since at
these energies we expect the behavior to be more normal-state
like \cite{feng1}. This is in particular true if the system is
close to a quantum critical point as was shown by Valla {\it et
al.} \cite{valla}.

Our results however are consistent with an underlying stripe
structure, as are, in a similar way, the ARPES data\cite{zhou1}.
If we adopt a quasi one-dimensional approach (i. e., stripe
structure) to describe the normal state, it is clear that since
continuous global symmetries cannot be spontaneously broken in
one dimension, even at $T=0$, the system has to exhibit
interchain coupling at the onset of order\cite{erica}. In that
context the system will retain some of its one dimensional
nature, and thus one should observe in STM studies of point
impurities in high-$T_{c}$ the signature of both Friedel-like
oscillations and a fixed CDW at the stripe wavevector.

Modulations which reflect the spectrum of elementary excitations
can however be distinguished from those related to incipient
order. In particular, the peak in the Fourier transform of the
LDOS produced by incipient order tends to be phase coherent, in
agreement with Fig.\ 7.  At the same time, features that are not
due to incipient order will either have a random phase (if they
originate from noise), or a phase that is strongly energy
dependent (if they originate from dispersing features). Thus, as
suggested in the previous section, in order to focus on those
aspects of the STM spectrum that reflect incipient order, one
should integrate over a small but finite range of energies about
$E=0$; non-dispersive, phase coherent features of the spectrum
will thus be accentuated, while dispersive, phase incoherent
features will tend to cancel\cite{kivelson} as was demonstrated
in Figs.\ 8 and 9. Moreover, the energy dependence presented in
Fig.\ 7 was obtained theoretically for models that include
striped-charge order \cite{subir,demler1,vojta}. All these models
rely on pinning of dynamic spin and charge density waves
fluctuations in a d-wave superconductor by local imperfections
which preserve spin-rotation invariance, such as impurities or
vortex cores. In particular, Podolsky {\it et al.} derived an
explicit expression for the Fourier component of the local
density of states at the ordering wavevector, and demonstrated
that various patterns of translational symmetry breaking exhibit
different energy dependence. Comparing to our data presented
earlier \cite{howald}, which is reproduced here in Fig.\ 7, they
argued that the data are consistent with weak translational
symmetry breaking that includes charge density waves and direct
dimerization. Indeed Fig.\ 4 in their paper reproduces remarkably
well our results.

Finally we note that in addition to this static striped density of
states structure that spans the entire surface, gap structure
inhomogeneities are observed in all our samples. The movement of
LDOS to the middle of the gap, around a particular energy, and its
origin from states just above the gap suggests that the density
modulation and superconductivity are inseparable. The
inhomogeneities in the superconducting gap structure should
therefore also be reflected in the striped density of states.
Indeed, the striping does show local variation as is shown in
Fig.\ 9, together with topological defects in the striped
structure visible in Fig.\ 4 \cite{kivfradkin}.

\section{CONCLUSION}

In this paper we studied the LDOS spectra of slightly overdoped
BSCCO single crystals. Based on our data we conclude that the
studied samples exhibit two types of modulated features in the
energy dependent LDOS. First, there are peaks at wavevectors
corresponding to four-period modulation of the LDOS along the Cu-O
bond direction, which we take as evidence for stripe order.
Second, there are features, primarily at shorter wavevectors,
which may to be energy dependent. These may be attributed to a
combination of elementary excitation (quasiparticle-like)
dispersion\cite{wanglee} and the strong inhomogeneities in the gap
structure previously observed by several groups
\cite{howald1,pan1,lang}. Finally, it is remarkable that in this
single set of experiments we see locally the d-wave
superconducting gap, the pseudogap, and metallic stripes,
suggesting that all reflect aspects of the same physics.\\

While this paper was being completed, Hoffman {\it et al.}
\cite{hoffman2} published higher resolution but qualitatively
similar data to that first published by Howald {\it et al.}
\cite{howald} and to the data presented in this paper. However,
in their paper they focus on the dispersive features of the data,
overlooking the existence of peaks at the stripe wavevector which
appear for all their reported energies.
\\

%%%%%%End of text

\noindent {\bf Acknowledgements:} We thank Subir Sachdev and
Steve Kivelson for many useful discussions. Work supported by Air
Force Office of Scientific Research.  The crystal growth was
supported by the U. S. Department of Energy under contracts No.
DE-FG03-99ER45773 and No. DE-AC03-76SF00515, by NSF CAREER Award
No. DMR-9985067, and by the A. P. Sloan Foundation.

%%%%%%%%%%%% figures
\begin{figure}
\includegraphics[width=0.95 \columnwidth]{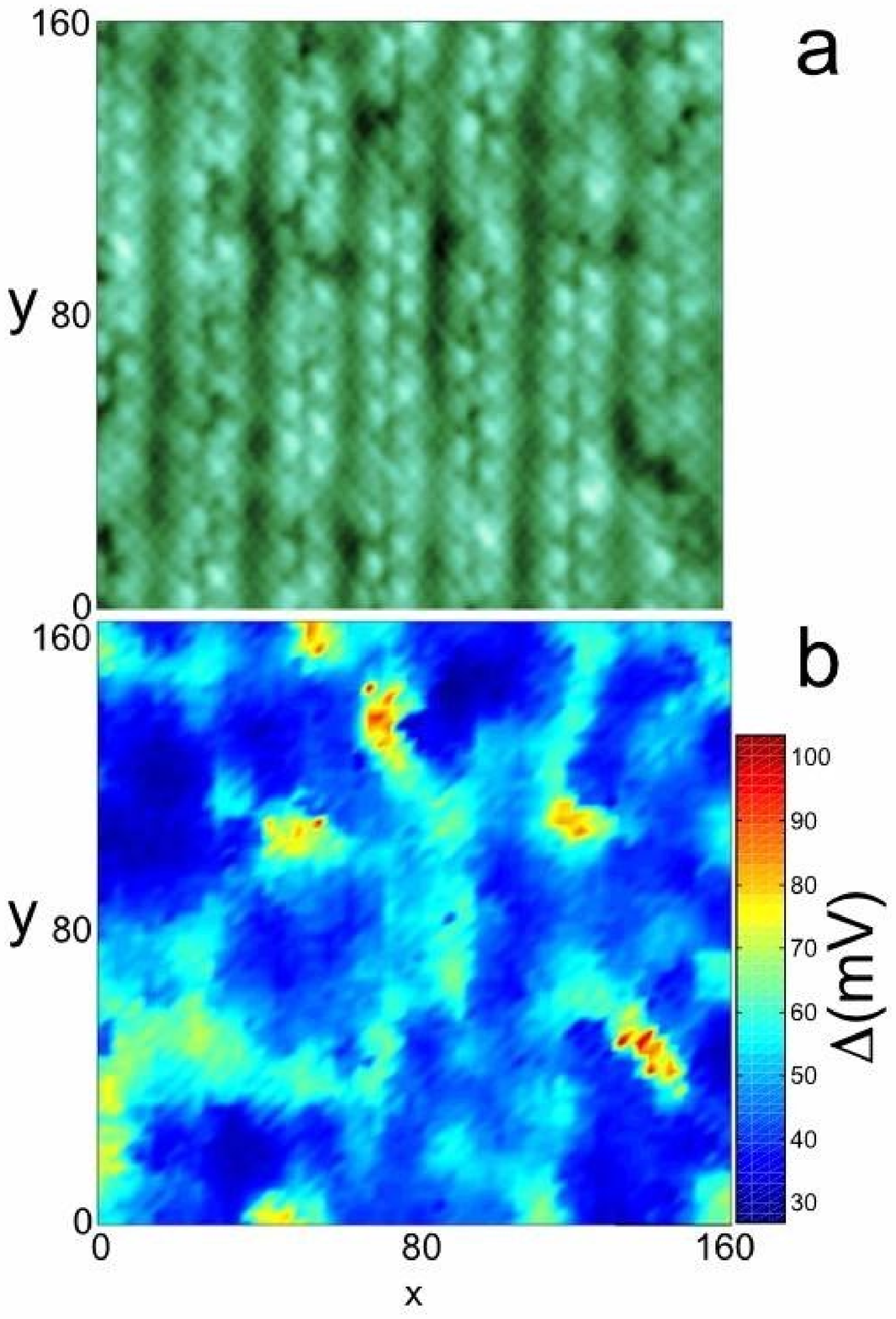}
\caption{Inhomogeneity in the electronic structure of a slightly
overdoped $\rm Bi_2Sr_2CaCu_2O_{8+\delta}$ single crystal. a)
Topography ($160$ \AA $\times 160$ \AA) of the cleaved BiO
surface. b) Gap size over the same area.}
 \label{fig1}
\end{figure}

\begin{figure}
\includegraphics[width=0.95 \columnwidth]{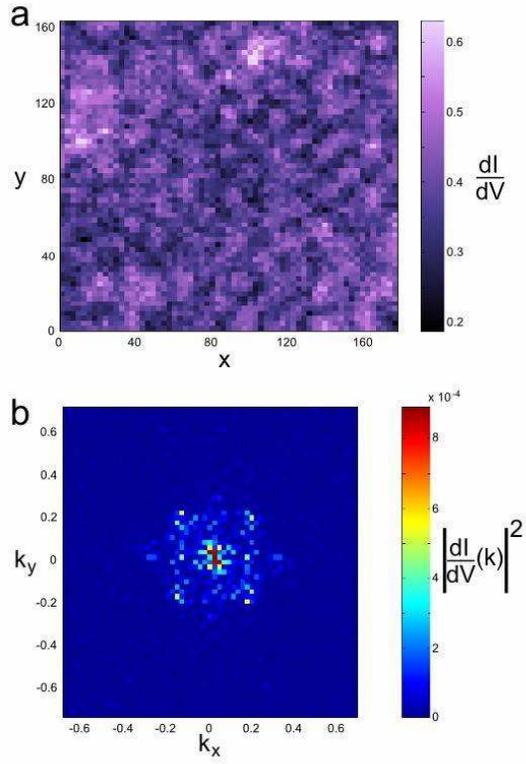}
\caption{Periodic spatial variation in the quasiparticle density
of states. a) Map of dI/dV(15 mV). b) Power spectrum of the
Fourier transform of dI/dV(15 mV).}
\label{fig2}
\end{figure}

\begin{figure}
\includegraphics[width=0.95 \columnwidth]{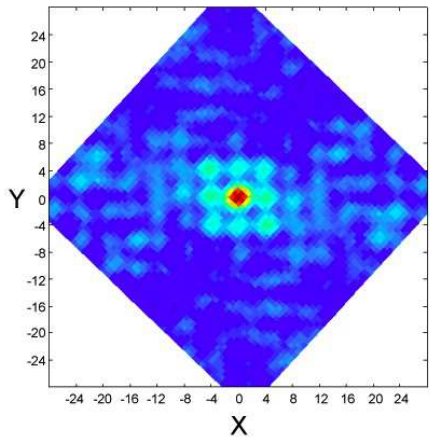}
\caption{Autocorrelation function of Fig.\ 2a showing the period
of the primary modulation (units are $a_0$).  Axes have been
rotated from Fig.\ 2a to make the Cu-O-Cu directions horizontal
and vertical.}
 \label{fig3}
\end{figure}

\begin{figure}
\includegraphics[width=0.95 \columnwidth]{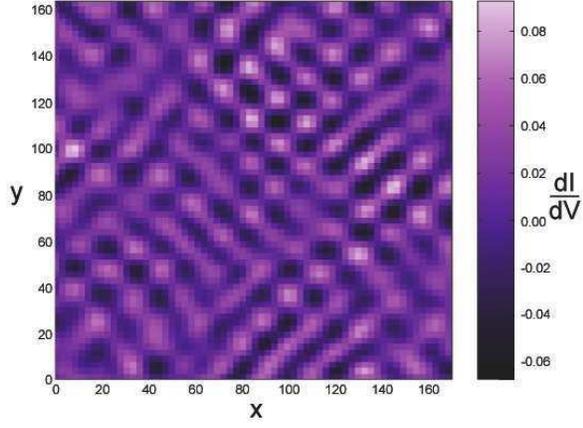}
\caption{Fourier filtered real-space map of dI/dV(15 mV) (Fig.
2a), using a sum of Gaussian $k$-space filters centered around
$k_{x,y}=\pm 2\pi / 4a_0$.}
\label{fig4}
\end{figure}

\begin{figure*}
\includegraphics[width=1.9 \columnwidth]{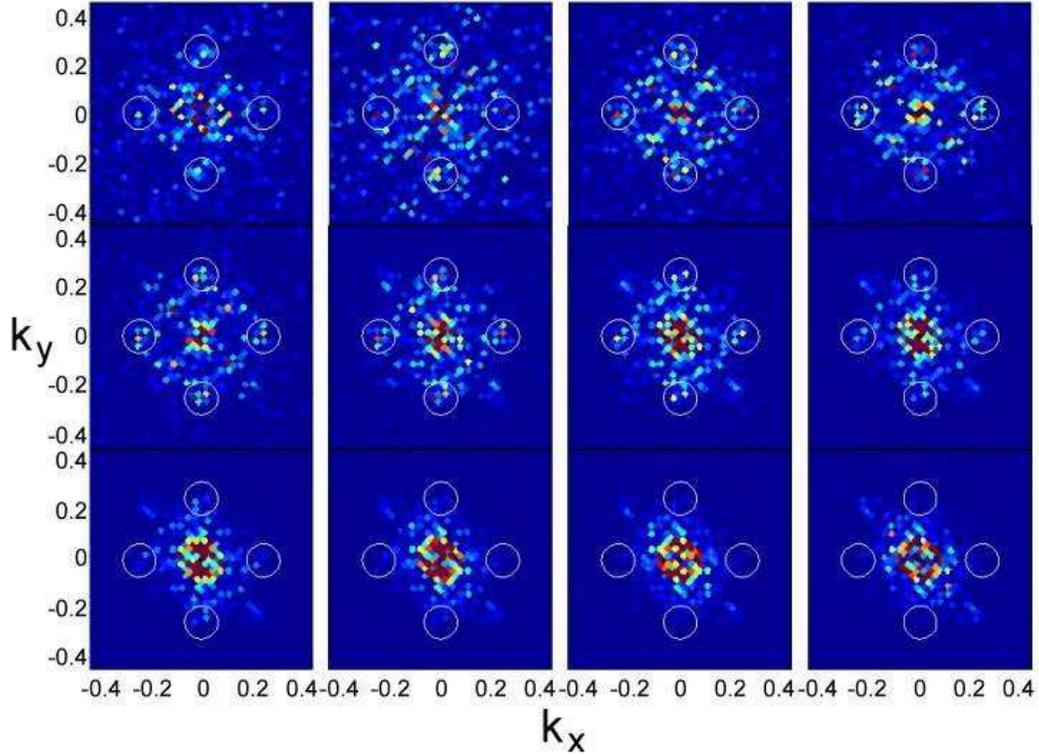}
\caption{Power spectra of dI/dV as a function of energy. Energies
are 0, 3, 7, and 10 mV (top row); 13, 16, 19, and 22 mV (middle
row); 25, 28, 32, and 35 mV (bottom row). White circles show the
width ($\sigma$) and position of the Gaussian filter used in Fig.
4. Axes have been rotated from Fig.\ 2.}
\label{fig5}
\end{figure*}

\begin{figure}
\includegraphics[width=0.95 \columnwidth]{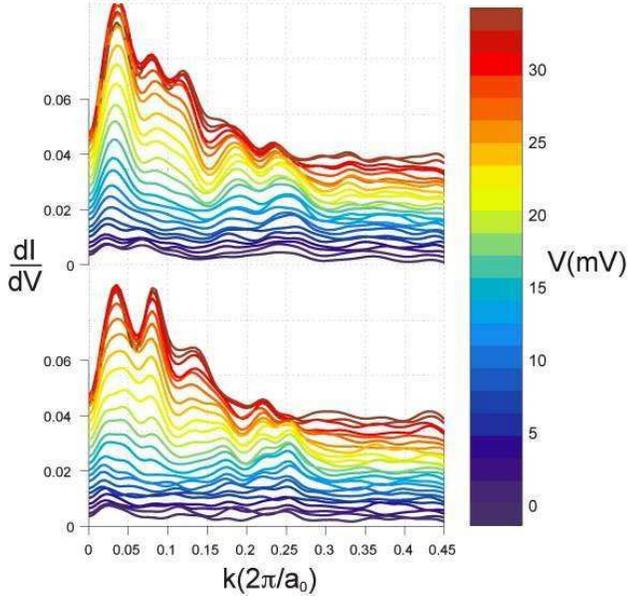}
\caption{Line scans along the (0,$\pi$) and ($\pi$,0) directions,
respectively, through each of the two independent Fourier
transform peaks as a function of energy ($|dI/dV(0,k_y,E)|$ and $|
dI/dV(k_x,0,E)|$). Scans are shifted vertically for clarity.}
\label{fig6}
\end{figure}

\begin{figure}
\includegraphics[width=0.95 \columnwidth]{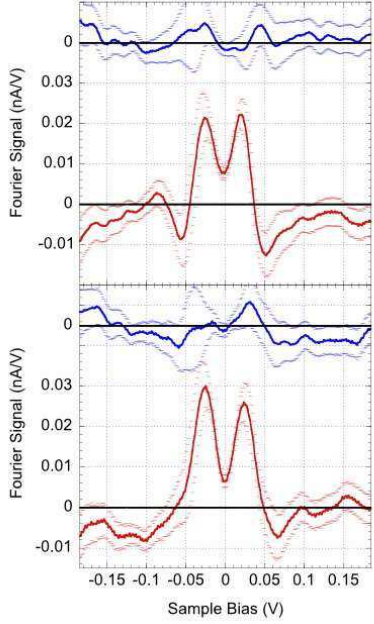}
\caption{Fourier transform at k=(2$\pi /4a_0$)($\pm$1,0) (top)
and k=(2$\pi /4a_0$)(0,$\pm$1) (bottom), the location of the
peaks in Fig.\ 2b, as a function of sample bias. The red and blue
traces correspond to the real and imaginary parts, respectively.}
\label{fig7}
\end{figure}

\begin{figure*}
\includegraphics[width=1.9 \columnwidth]{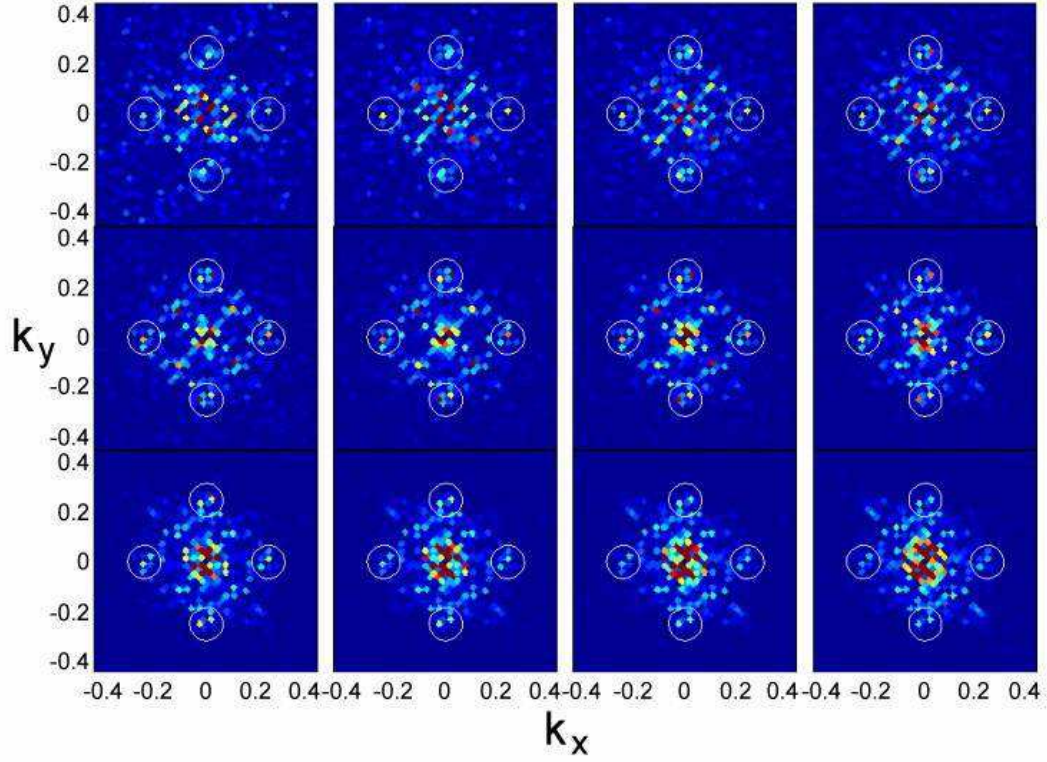}
\caption{Power spectra of I (integral of dI/dV) as a function of
energy. Energies used are the same as Fig. 5.}
 \label{fig8}
\end{figure*}

\begin{figure}
\includegraphics[width=0.95 \columnwidth]{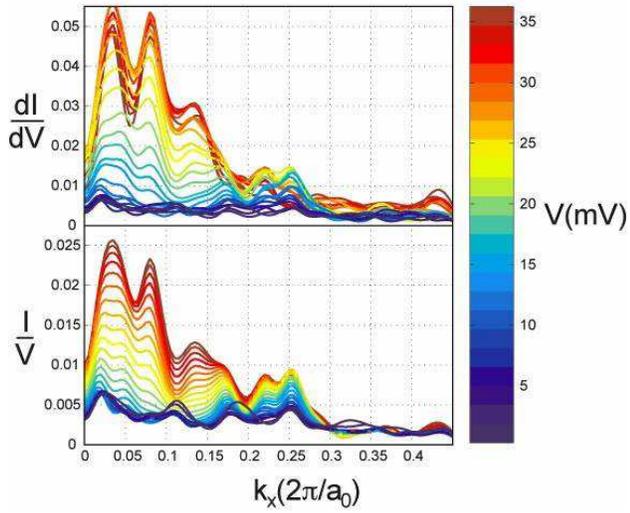}
\caption{Line scans as a function of $k_x$ along the $(0,0)$ to
$(\pi,0)$ direction, and as a function of energy (color scale).
Top panel shows the LDOS (dI/dV), and bottom panel the integrated
LDOS up to the given energy.}
 \label{fig9}
\end{figure}

\begin{figure}
\includegraphics[width=0.95 \columnwidth]{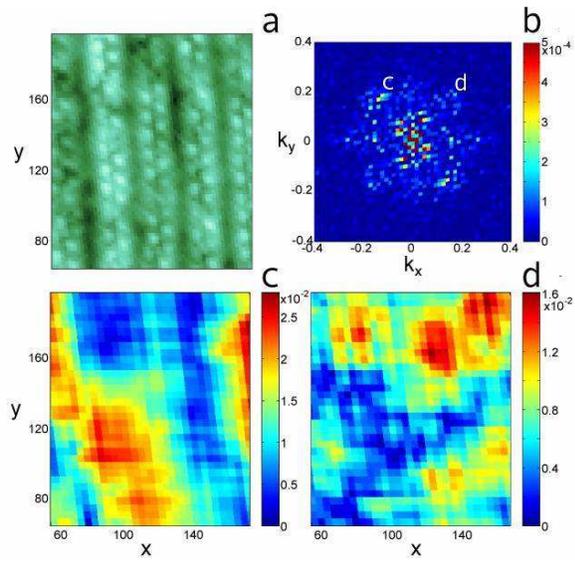}
\caption{Spatial variation in the amplitude of the periodic
modulation over a larger area. a) Constant current topograph ($130
$ \AA $\times 130$ \AA). b)  Power spectrum of dI/dV(15 mV) over
the entire $260$ \AA $\times 260$ \AA area showing the peaks used
in c) and d). c) Magnitude of the local Fourier transform of
dI/dV(15 mV) at one of the 2$\pi /4a_0$ points for a number of
$32\times 32$ pixel regions. d) The same map as c), for the other
peak.} \label{fig10}
\end{figure}

\end{document}